# Retrieving Dust Grain Sizes from Photopolarimetry: An Experimental Approach

O. Muñoz[1], E. Frattin[1,2], T. Jardiel[3], J. C. Gómez-Martín[1], F. Moreno[1], J. L. Ramos[1], D. Guirado[1], M. Peiteado[3], A. C. Caballero[3], J. Milli[4], and F. Ménard[4]  
[1] Instituto de Astrofísica de Andalucía, CSIC Glorieta de la Astronomía s/n, E-18008 Granada, Spain  
[2] Department of Physics and Astronomy G. Galilei, University of Padova, Vicolo dell Ósservatorio 3, I-35122 Padova, Italy  
[3] Instituto de Cerámica y Vidrio C/Kelsen 5, Campus Cantoblanco, E-28049 Madrid, Spain  
[4] Université Grenoble-Alpes, CNRS, IPAG F-38000 Grenoble, France

## Abstract

We present the experimental phase function, degree of linear polarization (DLP), and linear depolarization ($\delta_L$) curves of a set of forsterite samples representative of low-absorbing cosmic dust particles. The samples are prepared using state-of-the-art size-segregating techniques to obtain narrow size distributions spanning a broad range of the scattering size parameter domain. We conclude that the behavior of the phase function at the side- and back-scattering regions provides information on the size regime, the position and magnitude of the maximum of the DLP curve are strongly dependent on particle size, the negative polarization branch is mainly produced by particles with size parameters in the ∼6 to ∼20 range, and the $\delta_L$ is strongly dependent on particle size at all measured phase angles except for the exact backward direction. From a direct comparison of the experimental data with computations for spherical particles, it becomes clear that the use of the spherical model for simulating the phase function and DLP curves of irregular dust produces dramatic errors in the retrieved composition and size of the scattering particles: The experimental phase functions are reproduced by assuming unrealistically high values of the imaginary part of the refractive index. The spherical model does not reproduce the bell-shaped DLP curve of dust particles with sizes in the resonance and/or geometric optics size domain. Thus, the use of the Mie model for analyzing polarimetric observations might prevent locating dust particles with sizes of the order of or larger than the wavelength of the incident light.

*Unified Astronomy Thesaurus concepts:* Coma dust (2159); Spectropolarimetry (1973); Experimental techniques (2078); Debris disks (363)

## 1. Introduction

Dust is an important constituent in many astronomical environments. Dust particles scatter and absorb stellar radiation, affecting the radiative balance of the corresponding atmosphere. The angular dependence of the phase function, degree of linear polarization (DLP), and depolarization ratio ($\delta_L$) of the scattered light is dependent on the size, morphology, and refractive index of the scattering particles. Therefore, the analysis of the spectral dependence of the intensity and polarization of the scattered light is a powerful technique for characterizing cosmic dust particles.

Observations of the brightness and DLP of light scattered by comet dust have been conducted for a long time. Time variations of the coma and tail brightness as observed from Earth are not only dependent on the phase angle but also on the dust production rate as the comet moves in its orbit around the Sun. The DLP is a relative quantity that, in contrast to brightness, does not depend on the number of particles but on their physical properties (size, shape/structure, and composition). For that reason, spectropolarimetry is an essential tool for characterizing dust particles in cometary comae (see, e.g., Kiselev et al. 2015). Recently, the OSIRIS camera system on board the Rosetta mission (Sierks et al. 2015) has provided unique observations of the light scattered by dust within the coma of comet 67P/Churyumov–Gerasimenko (Bertini et al. 2017).

The depolarization ratio ($\delta_L$) obtained from lidar observations is widely used for characterizing atmospheric aerosols in Earth's atmosphere (e.g., Miffre et al. 2019; Kahnert et al. 2000). This quantity is not commonly used in astronomy although it has become of interest since the work of Sterzik et al. (2012). Those authors proposed using earthshine (the sunlight scattered by Earth and reflected from the lunar surface back to Earth) to observe our planet as an exoplanet. Such an approach can be helpful for detecting spectropolarimetric biosignatures in Earth-like exoplanets. In that context, the observed polarization spectra must be scaled to subtract the depolarizing effect of earthshine by reflecting on lunar regolith.

Regarding circumstellar regions, the arrival of high-contrast imaging observations broke new ground in characterizing grain size evolution by analyzing the scattered phase function and/or DLP curves (Canovas et al. 2015; Kataoka et al. 2015, 2017; Milli et al. 2017; Birnstiel et al. 2018; Ren et al. 2019; Arriaga et al. 2020).

The interpretation of the aforementioned observational data is hampered by the difficulty of the available light scattering codes in dealing with broad size distributions of realistic particle shapes. Great effort is being made to compute the scattering pattern of irregular cosmic dust particles (see, e.g., Min et al. 2003, 2010; Muinonen et al. 2009; Zubko et al. 2013; Merikallio et al. 2015; Pohl et al. 2016; Kolokolova et al. 2018; Moreno et al. 2018; Tazaki et al. 2016, 2019; Kirchslager & Bertrang 2020), even though computational simulations of the scattering pattern of cosmic dust grains are still constrained to certain size ranges and/or simplified shapes. Therefore, it is important to know the effects of the adopted particle shape model on the retrieved dust parameters such as dust grain and the refractive index. For instance, the simultaneous analysis of the observed phase function and DLP curves places emphasis on the limitations of the spherical particle model for characterizing real dust particles in debris disks (Graham et al. 2007; Arriaga et al. 2020; Engler et al. 2020). Further, the simultaneous analysis of the OSIRIS data set combined with ground-based DLP observations of comet 67P/Churyumov–Gerasimenko has exposed challenging contradictions in the retrieved physical properties of cometary dust (Hadamcik et al. 2016; Markkanen et al. 2018; Moreno et al. 2018; Muñoz et al. 2020). Understanding fundamental aspects of the interaction of electromagnetic radiation and dust particles is key to a breakthrough in the interpretation of photopolarimetric observations. Controlled scattering experiments with well-characterized natural dust samples provide key information to establish the link between dust physical properties and the way they scatter light in all directions.

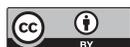

This work is part of an ongoing experimental project devoted to disentangling size, composition, and shape effects on the scattering behavior. The lack of control on the size distribution of top-down (grinding and sieving) approaches for synthesizing dust analog samples has hindered obtaining a relation between photopolarimetric features and the size of the scattering grains. First attempts (Muñoz et al. 2000; Volten et al. 2006) produced various size distributions out of bulk olivine powders. In those cases, the sieving procedure was not efficient enough for retrieving narrow size distributions. In this work the size distribution production relies on processing routines from the field of functional, nano-, and microceramics for synthesizing well-defined narrow size distributions. A low-absorbing bulk sample consisting of compact forsterite millimeter-sized grains has been processed to obtain five narrow size distributions spanning a wide scattering size parameter domain, namely: Rayleigh-resonance, resonance, resonance-geometric optics, and geometric optics. In the near future we expect to conduct a similar study with a set of samples consisting of highly absorbing particles to disentangle the effect of both size and absorption on the measured scattering pattern.

We present the experimental phase functions, DLP, and $\delta_L$ curves for the forsterite samples at 514 nm. The experimental data are obtained at the IAA-Cosmic Dust Laboratory (Muñoz et al. 2011). In Section 2 we present a brief description of the experimental apparatus and scattering-matrix formalism. The samples processing is conducted at the facilities of the Funceramics group at ICV-CSIC. Sample synthesis and characterization are provided in Section 3. The experimental data are presented in Section 4. We evaluate the performance of the Mie spherical model for reproducing the scattering pattern of natural dust grains in Section 5. Discussion and conclusions are summarized in Section 6.

## 2. Experimental Apparatus and Scattering-matrix Formalism

A detailed description of the scattering-matrix formalism, experimental apparatus, and data acquisition procedure is provided in Muñoz et al. (2010, 2011). In this work, we use as a light source a diode fiber pigtail laser that emits at 514 nm. We combine electro-optic modulation of the incident beam with lock-in detection to increase the accuracy of the measurements and concurrent determination of several elements of the 4 × 4 scattering matrix, **F** (Hovenier et al. 2004). The modulated laser beam is scattered by the cloud of particles under study located in a jet stream produced by an aerosol generator. The dust particles are brought into the jet stream as follows: The dust load is stored in a cylindrical reservoir with a piston that pushes the particles into the aerosol generator dispersion unit that contains a rotating steel brush. The dispersed powder is carried from the dispersion unit by a turbulent air stream to the nozzle placed above the scattering volume. The scattered light is detected by a photomultiplier tube, the Detector, that moves along a 1 m diameter ring covering the scattering angle range from 3° to 177°. Another photomultiplier tube, the Monitor, is located at a fixed position on the ring and is used to correct for potential fluctuations of the jet stream and/or the laser power.

The elements $F_{ij}$ of the scattering matrix are dimensionless and depend on the physical properties of the particles (morphology, size, and refractive index), wavelength of the incident beam, and the direction of scattering, i.e., the direction from the particle to the detector. The direction of scattering is defined by the angle between the directions of propagation of the incident and scattered beams, i.e., the scattering angle $(0 \leqslant \theta \leqslant \pi)$, and an azimuth angle, $\phi$, that ranges from 0 to $2\pi$. For randomly oriented particles, as is the case in our experiment, all scattering planes are equivalent and the scattering direction is fully described by means of the scattering angle. To facilitate direct comparison with astronomical observations, the scattering-matrix elements are presented in this work as functions of the phase angle, $\alpha = 180° - \theta$. If the cloud under study has a sufficient number of particles such that mirror symmetry in the particle ensemble can be safely assumed, we have (see, e.g., Hovenier et al. 2004, Section 2.4.1)

$$\begin{pmatrix} I_s \\ Q_s \\ U_s \\ V_s \end{pmatrix} \propto \begin{pmatrix} F_{11} & F_{12} & 0 & 0 \\ F_{12} & F_{22} & 0 & 0 \\ 0 & 0 & F_{33} & F_{34} \\ 0 & 0 & -F_{34} & F_{33} \end{pmatrix} \begin{pmatrix} I_0 \\ Q_0 \\ U_0 \\ V_0 \end{pmatrix}, \quad (1)$$

where $\{I_0, Q_0, U_0, V_0\}$ and $\{I_s, Q_s, U_s, V_s\}$ are the Stokes parameters that characterize the flux and the state of the linear and circular polarization of the incident and scattered beams, respectively (Hovenier et al. 2004). For natural incident light $(\{I_0, Q_0, U_0, V_0\} = \{1,0,0,0\})$, the first element of the scattering matrix, $F_{11}(\alpha)$, is proportional to the flux of the scattered light and is called the phase function. In that case, $U_s = 0$ and the $-(F_{12}(\alpha)/F_{11}(\alpha))$ ratio equals DLP, defined in terms of Stokes parameters as

$$\mathrm{DLP} = \frac{(Q_s^2 + U_s^2)^{1/2}}{I_s} = -\frac{F_{12}}{F_{11}}. \quad (2)$$

Similarly, if the incident beam is 100% linearly polarized parallel to the scattering plane $((\{I_0, Q_0, U_0, V_0\} = \{1,1,0,0\})$, the linear depolarization ratio $\delta_L$ is defined as

$$\delta_L = \frac{I - Q}{I + Q} = \frac{1 - F_{22}/F_{11}}{1 + 2F_{12}/F_{11} + F_{22}/F_{11}}. \quad (3)$$

In the case of a cloud consisting of spherical particles, the $F_{22}/F_{11}$ ratio equals unity at all phase angles. In this work, we present the measured $F_{11}(\alpha)$, $-(F_{12}(\alpha)/F_{11}(\alpha))$, $(F_{22}(\alpha)/F_{11}(\alpha))$, and corresponding $\delta_L(\alpha)$ of our set of forsterite samples. The lack of measurements at the forward and backward directions limits the use of the measured scattering-matrix elements for radiative transfer calculations. To facilitate the use of the data, we construct synthetic scattering-matrix elements from our measurements. The synthetic elements are defined in the full phase angle range from 0° to 180°. The extrapolation technique is defined in Escobar-Cerezo et al. (2018) and Gómez Martín et al. (2021). The extrapolated phase functions, $F_{11}^{\text{ext}}(\alpha)$, are normalized so that the following equation holds:

$$\frac{1}{2} \int_0^\pi d\theta \sin\theta F_{11}(\alpha) = 1. \qquad (4)$$

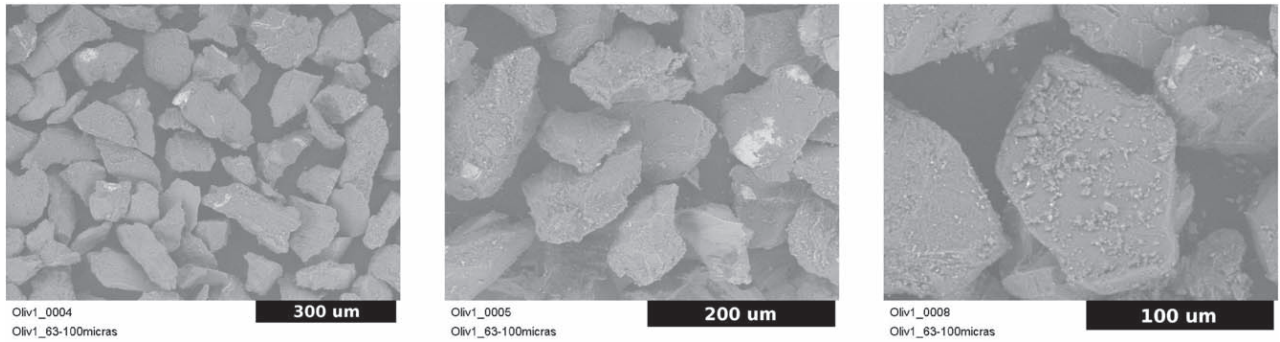

**Figure 1.** SEM images of the XL sample. The black bars at the bottom-right corner indicate 300 $\mu$m, 200 $\mu$m, and 100 $\mu$m in the left, middle, and right panels, respectively.

## 3. Sample Description

The initial forsterite powder shows two principal mineral phases via X-ray diffraction: $(Mg, Fe)_2SiO_4$ and $Mg_3Si_2O_5(OH)_4$. The sample has been processed to produce well-defined narrow size distributions. For this purpose, the bulk sample was first milled in a planetary mill for 4 hr with 0.3% of Dolapix CE64 dispersant (Zschimmer & Schwarz) to promote the dispersion of the particles. The milled powder was first sieved through a 100 $\mu$m mesh to discard any particle larger than that size; the obtained deposit was again sieved through a 63 $\mu$m mesh, yielding a first fraction of large particles labeled as sample XL (63 $\mu$m $\leqslant d \leqslant$ 100 $\mu$m). Figure 1 shows scanning electron microscope (SEM) images of the XL sample, which demonstrate the high efficiency of the sieving procedure. Still, a minor residual of submicron particles remains attached to the surfaces of large particles (Figure 1, right panel). Subsequently, the portion of the sample that passed through the 63 $\mu$m sieve was dispersed in ethylene glycol with a high-performance dispersion instrument (ultraturrax). The suspended particles were subjected to a 2 hr decanting process for removing the largest sizes: this first sediment is discarded and the supernatant is further separated by gravity in a glass vessel. Using this method, three different sediments were extracted after 8 hr, 15 hr, and 27 hr, respectively; the three of them were further processed by dispersion in ethylene glycol and subsequently centrifuged to remove submicron fragments attached to particle surfaces. As a result, three new samples were obtained: sample L (8 hr centrifuged deposit), sample M (15 hr centrifuged deposit), and sample S (27 hr centrifuged deposit). Field emission SEM (FESEM) images of the L, M, and S samples are displayed in Figure 2. These FESEM images demonstrate that the combination of dispersion and centrifugation is a highly efficient procedure for removing submicrometer particles from the surface of micron-sized particles. The supernatant of the 27 hr centrifuged deposit was dried in a heater at 60°, producing the powder fraction consisting of the smallest particles, namely sample XS. As shown in Figure 3, the particles of the XS sample tend to stick to each other as a result of electrostatic forces. In the course of the scattering measurements, the powder is dispersed in the aerosol generator dispersion unit before being delivered to the scattering volume to avoid the agglomeration of particles. Because the five forsterite samples have been produced by milling and sieving from the same original millimeter-sized grains, we do not expect significant differences in their particles shapes. Indeed, as shown in Figures 2 and 3, the five samples consist of compact particles with irregular shapes and sharp edges. Therefore, the effect of differences in shape on the scattering behavior is not taken into account in the discussion section.

The size distributions for the XL, L, M, S, and XS samples were obtained with a laser light scattering particle sizer (Malvern Mastersizer 2000). The particle sizer measures the phase function of the sample dispersed in a carrier fluid at 633 nm. The measured phase function spans a scattering angle range between 0°.02 and 135°, with special attention to the forward scattering peak. The technique is based on the assumption that the forward scattering peak for randomly oriented particles with moderate aspect ratios mainly depends on particle size and is weakly dependent on particle shape (Mishchenko et al. 1996, 1997). The volume distribution of equivalent spherical particles is then retrieved by inverting the observed phase function using a light scattering kernel based on Mie theory. In the size distribution retrieval, we assume a refractive index ($m = n + ik$) for the forsterite samples $m = 1.65 + i10^{-5}$ (Huffman & Stapp 1973). The Mie model constrains the intrinsic applicability method to $r > 0.1$ $\mu$m (Gómez Martín et al. 2020). Therefore, the size distribution of the XS sample is retrieved by combining data from Mastersizer and Zetasizer. The Zetasizer retrieval is based on the dynamic light-scattering technique. It determines the size by measuring the Brownian motion of submicron particles dispersed in a liquid.

In Figure 4 (left panel), we show the retrieved projected surface equivalent distributions, $S(\log r)$. $S(\log r)d(\log r)$ gives the relative contribution of spheres in the size range [$\log r$; $\log r + d\log r$] to the total projected surface per unit volume. Taking into account the monomodal nature of the five size distributions, they can be safely characterized by their effective radius, $r_{\text{eff}}$, and variances, $v_{\text{eff}}$, defined as (Hansen & Travis 1974):

$$r_{\text{eff}} = \frac{\int_0^\infty r \pi r^2 n(r) dr}{\int_0^\infty \pi r^2 n(r) dr}, \qquad (5)$$

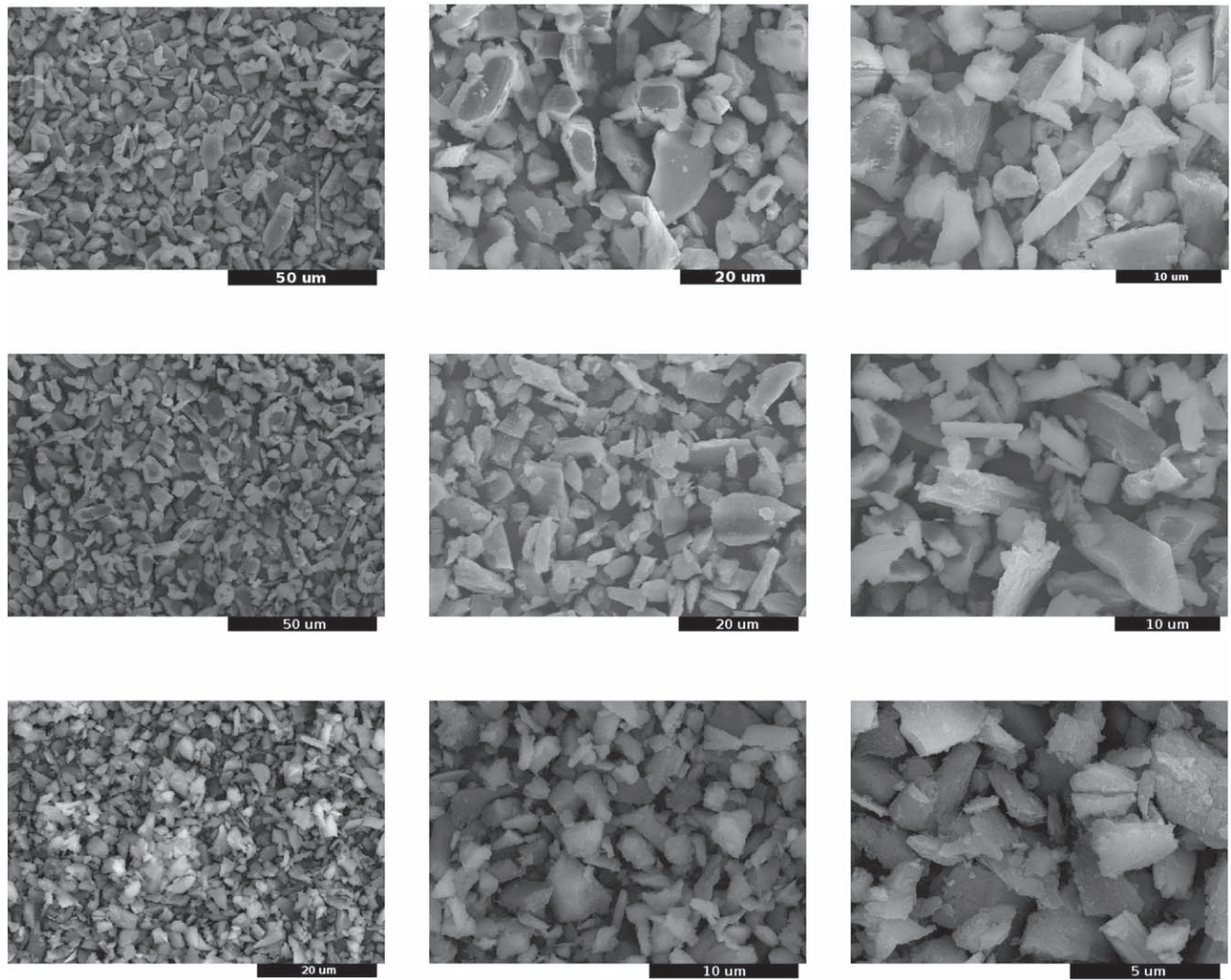

**Figure 2.** FESEM images of the L (top), M (medium), and S (bottom) samples. In the first and second rows, black bars at the bottom-right corner of the images indicate 50 μm, 20 μm, and 10 μm in the left, middle, and right panels, respectively. In the third row the black bars denote 20 μm, 10 μm, and 5 μm in the left, middle, and right panels, respectively.

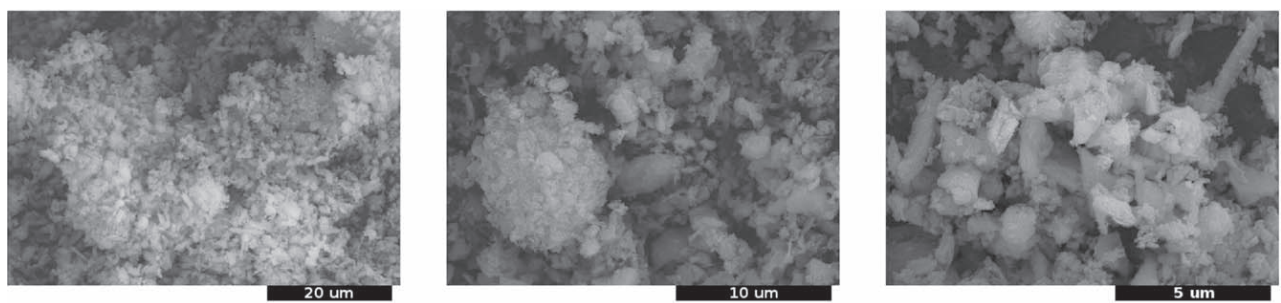

**Figure 3.** FESEM images of the XS sample. The black bars at the bottom-right corners of each panel denotes 20 μm, 10 μm, and 5 μm in the left, middle, and right panels, respectively.

$$v_{\rm eff} = \frac{\int_0^\infty (r - r_{\rm eff})^2 \pi r^2 n(r) dr}{r_{\rm eff}^2 \int_0^\infty \pi r^2 n(r) dr}, \quad (6)$$

where $n(r)dr$ is the fraction of the total number of projected surface equivalent spheres with radii in the size range $[r, r+dr]$ per unit volume of space. Here, $n(r)$ is computed from the measured volume distributions (Figure 4, right panel). Table 1 lists the values $r_{\rm eff}$ and $v_{\rm eff}$ and the corresponding effective size parameter $x_{\rm eff}$ at the experiment wavelength of 514 nm ($x_{\rm eff} = 2\pi r_{\rm eff}/\lambda$). The effective size parameter range spans from 4 for the XS sample to 575 for the XL sample. With regard to the size of the scatterers, the scattering regimes are defined as: (1) Rayleigh scattering ($|mx| \ll 1$, where $m$ is the complex refractive index), e.g., that produced from gas molecules in the visible; (2) Rayleigh-resonance transition regime ($\lesssim \lambda$) also known as Rayleigh–Mie; (3) resonance (Mie) regime ($r \approx \lambda$); and (4) geometric optics regime ($r \gg \lambda$). Note

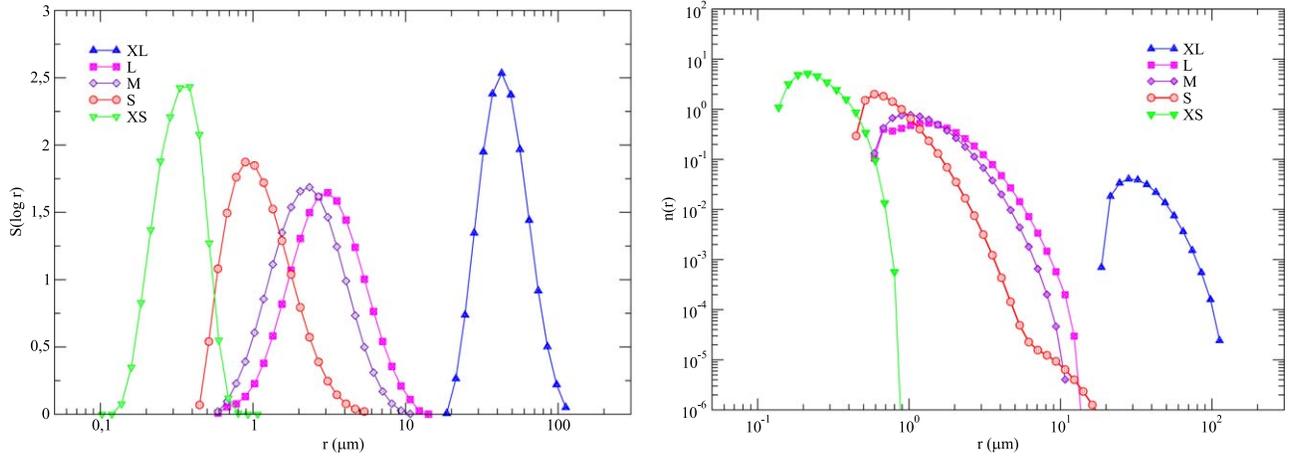

**Figure 4.** Projected surface area distributions (left) and number distributions (right) of the XL (blue triangles up), L (magenta squares), M (purple diamonds), S (red circles), and XS (green upside-down triangles) as functions of radius in micrometers on a logarithmic scale.

**Table 1**
Measured Main Parameters of the Phase Function Curves for the Five Size-segregated Samples and Quartzite Pebble

| Sample | $r_{\rm eff}$ ($\mu$m) | $v_{\rm eff}$ | $x_{\rm eff}$ | $f$[a] | BS[b] |
|---|---|---|---|---|---|
| XS | 0.36 | 0.42 | 4 | 0.71 | 1.57 |
| S | 1.4 | 1.06 | 17 | 0.63 | 1.52 |
| M | 2.6 | 0.55 | 32 | 0.71 | 1.50 |
| L | 3.5 | 0.30 | 43 | 0.80 | 1.30 |
| XL | 47 | 0.12 | 575 | 0.84 | 1.42 |
| Pebble | 1.6[c] | ⋯ | ∼200,000 | 1.33 | 1.19[d] |

**Notes.**
[a] $f = F_{11}(45°)/F_{11}(90°)$.
[b] BS $= F_{11}(0°)/F_{11}(45°)$.
[c] $r$(mm).
[d] BS value computed with the measured smallest phase angle, 10°.

that, strictly speaking, Mie scattering refers to scattering by homogeneous spherical particles. When dealing with irregular particles, the Mie regime refers to the particle size that is of the order of the wavelength of the incident light. To avoid confusion, we refer to the resonance scattering regime throughout this paper when dealing with sizes of particles of the order of $\lambda$. Therefore, our set of forsterite samples allows us to study the effect of size on the measured scattering-matrix elements spanning over nearly the full scattering size parameter domain: Sample XS consists of particles with sizes in the transition region between the Rayleigh and resonance scattering regimes; samples S, M, and L, belong to the resonance and/or transition region between the resonance and geometric optics regimes; and sample XL consists of particles in the geometric optics regime.

## 4. Measurements

### 4.1. Phase Functions

Figure 5 shows the extrapolated phase function curves, $F_{11}^{ext}(\alpha)$, for the forsterite XL, L, M, S, and XS samples. All measured phase functions show strong forward peaks and a rather flat dependence on the phase angle at the side- and back-scattering regions. The forward scattering lobe (region A in Figure 5) is strongly dependent on grain size. The two samples consisting of larger particles (XL and L) show the narrowest forward peaks. Further, in spite of the apparent flat side- and back-scattering trends, some differences between the five samples can still be observed. The flattening of the $F_{11}(\alpha)$ curves at side-phase angles (region B in Figure 5) is evaluated by means of the estimator $f$ defined as $f = F_{11}(45°)/F_{11}(90°)$. The $f$ values corresponding to the five samples are reported in Table 1. The closer the $f$ value is to 1, the flatter the curve. We see that the flattest curves correspond to the samples consisting of larger particles (XL and L). The value of $f$ decreases when the size of the particles decreases (samples M and S), slightly increasing again for sample XS. The estimator BS $= F_{11}(0°)/F_{11}(45°)$ in Table 1 evaluates the behavior of the experimental phase functions at the backward direction (region C in Figure 5). Again, the flattest behavior is presented by the samples consisting of larger particles (XL, and L). BS deviates more strongly from unity when approaching the Rayleigh scattering regime (sample XS). The increase in the scattered flux in the backward region is typical for nonabsorbing particles smaller than the wavelength (see, e.g., Liu et al. 2015). In the limiting case, pure Rayleigh scattering ($r \ll \lambda$), the phase function shows a symmetrical dependence on the phase angle with maxima at 0° and 180° and a minimum at 90°. In the opposite limiting case of particles much larger than the wavelength ($r \gg \lambda$) the phase function shows significant differences at the side- and back-scattering regions. Figure 5 also displays the experimental phase function of a millimeter-sized ($r = 1.6$ mm) quartzite pebble (Muñoz et al. 2020). The refractive index of quartzite ($m = 1.58+i2\ 10^{-5}$) is similar to that of the forsterite samples. However, its phase function is significantly different. The measured $F_{11}(\alpha)$ for the quartzite pebble is monotonically decreasing from 10° to 160°. Indeed, the $f$ estimator of the flattening of the phase function for the pebble is >1, showing an opposite slope to that of the micron-sized forsterite samples. The diffraction peak of such a large particle is located at ±0°.8 around the exact forward direction (180°), which is outside of the measured range (10°–170°).

### 4.2. Degree of Linear Polarization

The measured DLP curves in percentage for the XL, L, M, S, and XS samples are plotted in Figure 6. To facilitate comparison between different size regimes, red and green dashed horizontal lines represent the measured maximum and minimum DLP values of the complete set of measurements, respectively. All measured DLP curves are bell shaped, with maxima at side-phase angles and negative polarization branches (NPB) at small phase angles. Table 2 lists the main parameters of the DLP curves in the region of minimum (DLP$_{min}$, $\alpha_{min}$) and maximum (DLP$_{max}$, $\alpha_{max}$) polarization and polarimetric slope ($h$) at the inversion angle ($\alpha_0$). In Figure 6 (bottom-right panel) we plot the measured DLP curve of the millimeter-sized quartzite pebble.

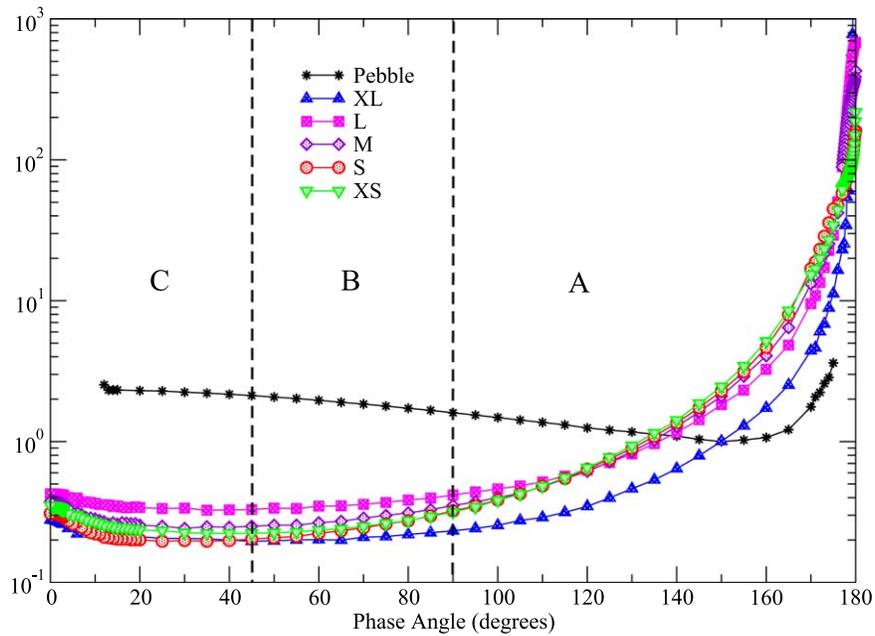

**Figure 5.** Extrapolated phase functions ($F_{11}^{ext}(\theta)$) curves at 514 nm for forsterite XL (triangle), L (squares), M (diamonds), S (circles), and XS (upside-down triangle). Black stars correspond to the experimental $F_{11}(\theta)$ curve of a quartzite pebble ($r = 1.6$ mm) after Muñoz et al. (2020). The experimental phase function for quartzite is arbitrarily normalized to unity at 150°.

Figure 6 shows different trends for the maximum of the DLP curve depending on the size regime. The maximum DLP$_{max}$ (20.3%) is obtained for the sample consisting of particles with sizes in the Rayleigh-resonance transition regime, i.e., sample XS. As the size of the particles increases into the resonance regime, the value of DLP$_{max}$ decreases from 12.9% for sample S to 9.4% for sample L. The tendency of a decrease DLP$_{max}$ with the size parameter is reversed as we approach the geometric optics regime, obtaining a value of 15.9% for sample XL and increasing up to 20.3% for the millimeter-sized quartzite pebble (Muñoz et al. 2020). As the size of the scattering particles increases into the geometric optics regime, the position of DLP$_{max}$ shifts toward larger phase angles (see also Liu et al. 2015; Escobar-Cerezo et al. 2018).

The deepest negative polarization branch (DLP$_{min}$ = −4.4%) is obtained for sample S, which consists of particles with radii of the order of 1 $\mu$m ($r_{eff}$ = 1.4 $\mu$m; $x_{eff}$ = 17). |DLP$_{min}$| decreases as the particle size increases, showing values of 2.6% and 1.7% for samples M and L, respectively. In the case of the XL sample, the NPB vanishes within the error bars. Another interesting effect observed for the XS sample is that even though the shape of its NPB is very similar to that of sample S, it presents a high dispersion of values at the measured phase angle range (3°–20°, in steps of 1°) as shown by the larger error bars. This can be understood in terms of a mixture of particles with sizes in the resonance and Rayleigh scattering size regimes, where the NPB tends to disappear. All in all, for low-absorbing particles, the NPB is mainly produced by grains with radii in the ∼0.5 to 2 $\mu$m range (size parameters spanning from ∼6 to ∼20). To our knowledge, these measurements constitute the first experimental evidence for a strong dependence of the position and magnitude of the NPB on the grain size range. Previous DLP curves for compact millimeter-sized particles obtained at 514 nm also show shallow NPBs, with the position of their inversion angles dependent on the particle composition (Muñoz et al. 2020). For such large grains, micron-sized surface roughness and/or internal structures are likely responsible for the measured negative branch.

### 4.3. Linear Depolarization Ratio

In Figure 7 we present the measured $F_{22}(\alpha)/F_{11}(\alpha)$ curves (left panel) and corresponding linear depolarization ratios (middle panel) as defined in Equation (3). Due to the limited amount of sample, we could not measure this ratio for sample M. All samples studied show a decrease of $F_{22}(\alpha)/F_{11}(\alpha)$ from ∼1 at large phase angles to a minimum at side-phase angles, increasing again at the backward direction. The $F_{22}(\alpha)/F_{11}(\alpha)$ ratio shows a strong dependence on grain size. The samples consisting of smaller particles, XS and S, show the highest values across the complete measured phase angle range. Further, the behavior of the $F_{22}(\alpha)/F_{11}(\alpha)$ ratios for samples XS and S is nearly identical. The lowest values occur for sample XL at all measured phase angles. Our data show a correlation between the NPB and the $F_{22}/F_{11}$ ratio at the near-backward direction. This correlation is illustrated in Figure 7 (right panel), where the measured $-F_{12}/F_{11}$ curves are plotted alongside the $F_{22}/F_{11}$ curves for samples S and L. As shown, for sample S there is an anticorrelation between the NPB and a depolarization surge (DS) at the near-backward direction (position marked by a dashed line in Figure 7, right panel). As the NPB becomes shallower (sample L), the DS vanishes.

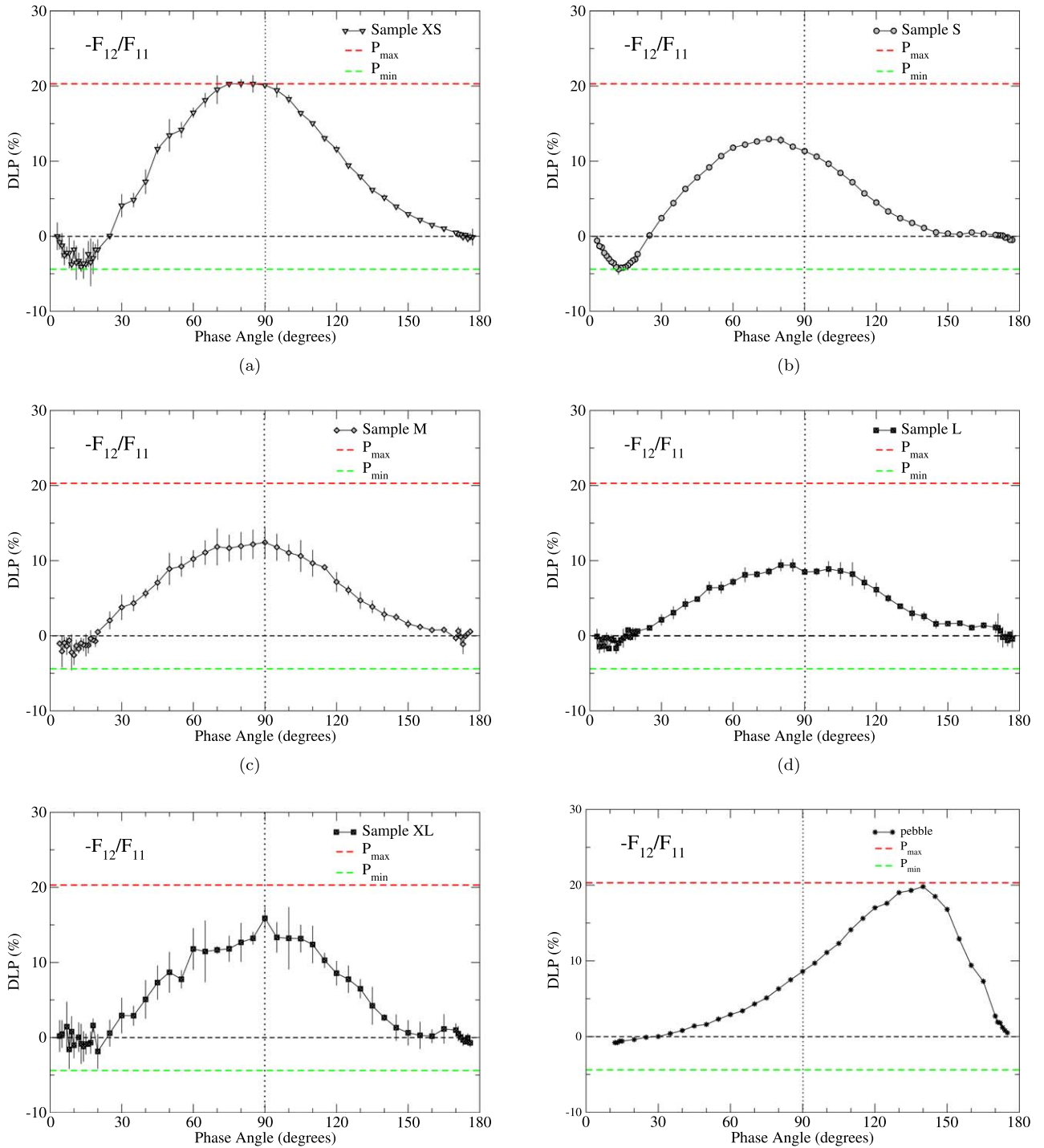

**Figure 6.** Measured degree of linear polarization curves in percentage as functions of the phase angle for samples XS and S (top panels), M and L (middle panels), and XL and quartzite pebble (bottom panels). The measured curves are presented with their error bars. In cases in which no error bars are shown, they are smaller than the symbol. Red and green dashed lines indicate the measured maximum and minimum DLP values, respectively.

## 5. Should We Use Mie Theory for Interpreting Photopolarimetric Observations of Dust Clouds?

Mie theory for homogeneous spherical particles has been (and still is) widely used for interpreting photopolarimetric astronomical observations of dusty objects. Computations of scattering and absorption properties of spherical particles can be accurately performed without any limitation on size and/or refractive index. However, as it is already well known in the terrestrial aerosols community, assuming the spherical model for irregularly shaped dust particles can produce dramatic errors in the computed scattering pattern (see, e.g., Mishchenko et al. 2003; Herman et al. 2005; Dubovik et al. 2006). The measured $F_{11}(\alpha)$ curves for the samples belonging to the Rayleigh-resonance (XS), resonance (S), and geometric optics (XL) scattering regimes are plotted in Figure 8, respectively. Green solid lines correspond to Mie computations for homogeneous spherical particles with the size distribution and refractive index of the corresponding forsterite samples at

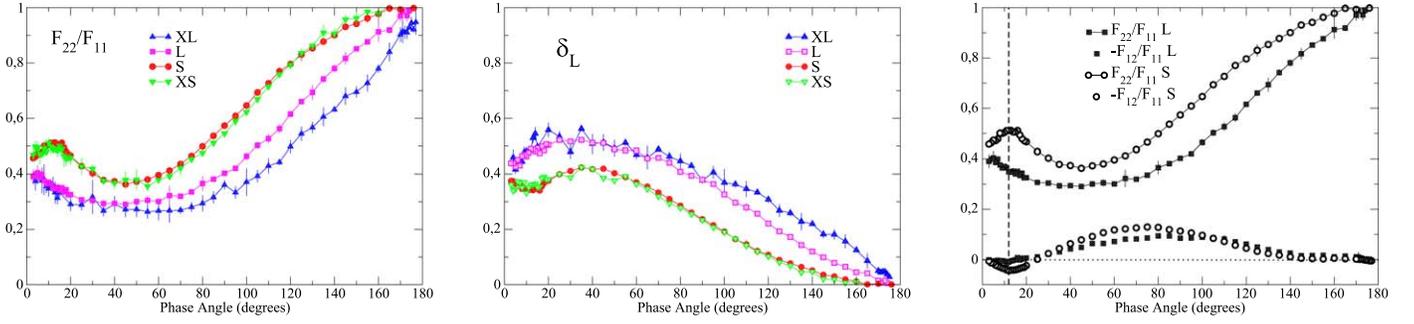

**Figure 7.** Measured $F_{22}/F_{11}$ ratios (left) and corresponding linear depolarization ratio (middle) for samples XL (triangles), L (squares), S (circles), and XS (upside-down triangles). The right panel shows the measured $-F_{12}/F_{11}$ and $F_{22}/F_{11}$ curves for samples L (squares) and S (circles), respectively.

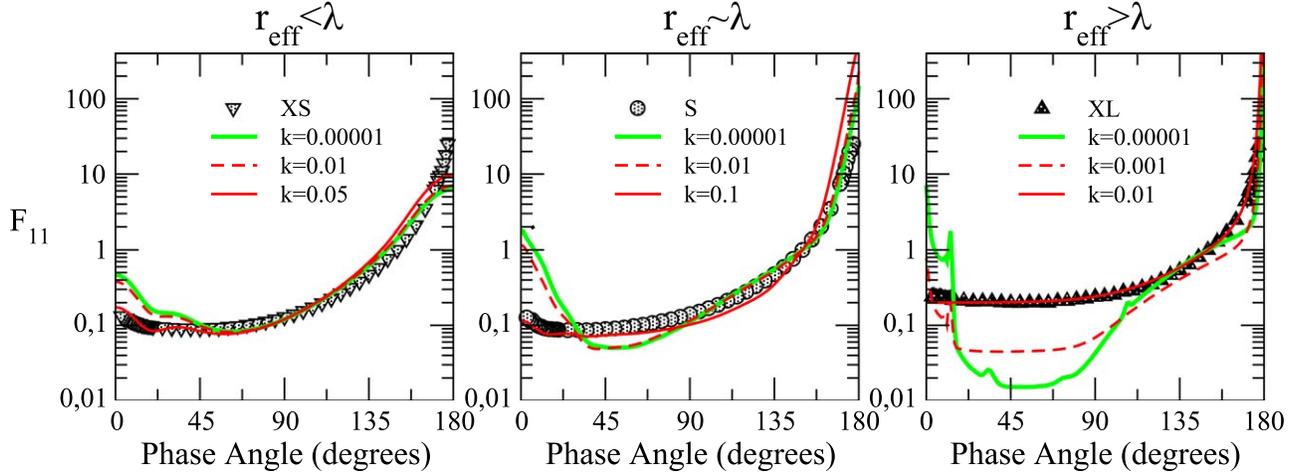

**Figure 8.** Experimental phase function curves for samples XS (left), S (middle), and XL (right). The experimental data are plotted alongside Mie computations for the corresponding size distributions. For the calculations, the real part of the refractive index is fixed to 1.65 and the imaginary part, $k$, is varied between 0.00001 and 0.003 (left), 0.1 (middle), and 0.01 (right).

Table 2
Measured Main Parameters of the Degree of Linear Polarization Curves

| Sample | $r_{eff}$ ($\mu$m) | $v_{eff}$ | $x_{eff}$ | DLP$_{min}$ (%) | $\alpha_{min}$ (deg) | $\alpha_0$ (deg) | DLP$_{max}$ (%) | $\alpha_{max}$ (deg) | $h$ (%/deg) |
|---|---|---|---|---|---|---|---|---|---|
| XS | 0.36 | 0.42 | 4 | $-4.1 \pm 0.6$ | $13 \pm 1$ | $25 \pm 1$ | $20.3 \pm 0.6$ | $80 \pm 5$ | 0.53 |
| S | 1.4 | 1.06 | 17 | $-4.4 \pm 0.7$ | $12 \pm 1$ | $25 \pm 1$ | $12.9 \pm 0.4$ | $75 \pm 5$ | 0.62 |
| M | 2.6 | 0.55 | 32 | $-2.6 \pm 1.3$ | $10 \pm 1$ | $20 \pm 1$ | $12.4 \pm 2.1$ | $90 \pm 5$ | 0.37 |
| L | 3.5 | 0.30 | 43 | $-1.7 \pm 0.3$ | $8 \pm 1$ | $15 \pm 1$ | $9.4 \pm 0.8$ | $80 \pm 5$ | 0.15 |
| XL | 47 | 0.12 | 575 | $-1.2 \pm 2.0$ | $14 \pm 1$ | $17 \pm 1$ | $15.9 \pm 1.5$ | $90 \pm 5$ | 0.28 |
| Pebble[a] | 1.6[b] | ... | ~200,000 | $-0.8$ | $12 \pm 1$ | $30 \pm 1$ | 19.8 | $140 \pm 5$ | 0.06 |

**Notes.**
[a] Data from Muñoz et al. (2020).
[b] $r$ (mm).

the measured wavelength ($1.65 + i10^{-5}$). Samples S and XL show strong differences between measured and computed values at the side- and back-scattering regions. The effect of particle shape on the phase function curve diminishes as the size of the particles relative to the wavelength decreases. As shown in Figure 8, left panel, the differences between the phase functions for spherical and irregular particles are smaller in the case of the sample consisting of a mixture of particles in the Rayleigh and resonance scattering size regimes. The differences in the phase function between spherical and nonspherical particles also decrease with increasing absorption

(Mishchenko et al. 1997). In Figure 8, solid and dashed red lines correspond to Mie computations for unrealistically high imaginary parts, $k$, of the refractive index. As shown, the computed phase function at the side- and back-scattering regions is flattened as the absorption is increased. Indeed, experimental and computed curves agree at nearly all phase angles for $k$ equal to 0.05, 0.1, and 0.01 for the XS, S, and XL samples, respectively. That is, the measured phase function for low-absorbing irregular forsterite particles can be reasonably well reproduced by highly absorbing spherical particles. Difficulties arise when retrieving the composition of the dust particles by using the spherical model. The observed phase function at side- and back-scattering regions can be reproduced by assuming unrealistically high values of the imaginary part of the refractive index and therefore erroneous information on the composition for the dust particles populating the cloud would be retrieved.

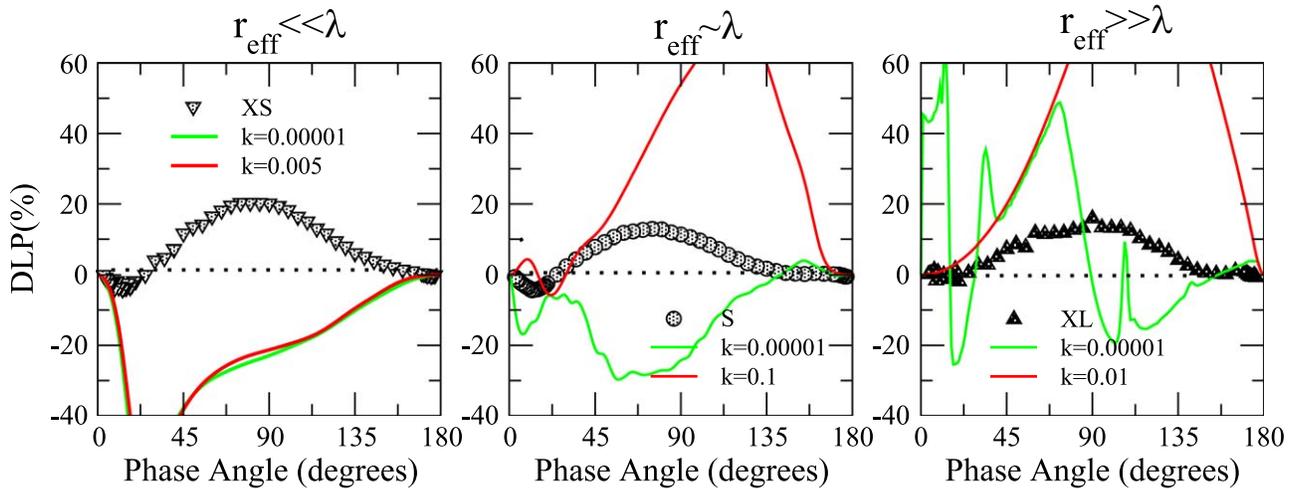

**Figure 9.** Experimental DLP ($-F_{12}(\alpha)/F_{11}(\alpha)$) curves in percentage for sample XS (left), S (middle), and XL (right). The experimental data are presented together with results of Mie computations for the corresponding size distributions. For the calculations, the real part of the refractive index, $n$, is fixed to 1.65 and the imaginary part, $k$, is varied between 0.00001 and 0.003 (left), 0.1 (middle), and 0.01 (right).

A similar test has been done with the DLP curves. Figure 9 shows the measured $-F_{12}(\alpha)/F_{11}(\alpha)$ curves for samples XS, S, and XL. As in the previous case, green solid lines correspond to Mie computations for the size distribution and refractive index of the corresponding forsterite sample at 514 nm. Differences in the DLP curves for spherical/irregular particles for samples XS, S, and XL are even stronger than those for the $F_{11}(\alpha)$ element. In contrast to the phase function test, the computed DLP curves do not approach the measured values when absorption is increasing (red lines). Interestingly, in the case of the sample belonging to the resonance regime, sample S, the sign of the computed DLP curve is even reversed when the imaginary part is increased from 0.00001 to 0.1.

In a second test, the size distribution parameters and refractive index are chosen so that differences between the measured and calculated values for the DLP curve are minimized. For the Mie calculations, we assume a power-law size distribution $n(r) \propto r^{-q}$. In the fitting procedure the parameters that define the size distribution, namely, the power-law exponent, $q$, the minimum, $r_{\min}$, and maximum radii, $r_{\max}$, and the real $n$ and imaginary $k$ parts of the refractive index, are considered as free parameters. The method to find the best-fit values is based on the downhill simplex method by Nelder & Mead (1965). We have used the FORTRAN implementation as described by Press et al. (1992). The initial set of parameters (the starting simplex) that feeds the iterative fitting process comprise a broad range of sizes and compositions. Table 3 lists the initial ranges of fitting parameters together with the best-fit parameters and actual values for the XS, S, and XL samples. Figure 10 shows the experimental and best-fit DLP curves for samples XS (left), S (middle), and XL (right). As shown, computed DLP curves for the best-fit parameters (dashed red lines) qualitatively reproduce the experimental curves. However, the retrieved size distributions are far from the actual size distribution of samples XS, S, and XL. For the three polarization curves, the spherical model retrieves effective size parameter values $x_{\mathrm{eff}} < 1.5$, indicating the presence of particles in the Rayleigh-resonance regime. However, each sample belongs to a different size regime: $x_{\mathrm{eff}}(\mathrm{XS}) = 4$, $x_{\mathrm{eff}}(\mathrm{S}) = 17$, and $x_{\mathrm{eff}}(\mathrm{XL}) = 575$. This shows that assuming a spherical model for interpreting polarimetric observations will result in significant errors in the size range of the retrieved grains. Regarding refractive indexes, the model also yields values very far from the actual one. In particular, the imaginary part retrieved from the Mie fitting is orders of magnitude higher than the actual values resulting in unrealistically high-absorbing particles.

## 6. Discussion and Conclusions

Simultaneous analysis of photometric and polarimetric observations of light scattered by dust grains at various wavelengths provide important constraints on the physical properties of the scattering particles. The use of an adequate model particle is mandatory for the proper interpretation of remotely observed photopolarimetric observations (see, e.g., Arriaga et al. 2020; Calcino et al. 2020; Duchêne et al. 2020). As shown in the previous section, the use of the spherical particle model for analyzing the light scattered by a cloud of irregularly shaped nonabsorbing dust particles produces significant errors in the retrieved grain parameters. The observed phase function at side- and back-scattering regions can be reproduced by assuming unrealistically high values of the imaginary part of the refractive index and therefore erroneous information on the composition of the dust particles populating the cloud. This could explain the unrealistic refractive index obtained by Duchêne et al. (2020) for the grain particles populating the debris disk HD 32297HD. The DLP curve is even more dependent on particle shape. The spherical model does not reproduce the bell-shaped polarization curve typical of dust particles with sizes in the resonance and/or geometric optics regime. Therefore, the use of the spherical model for analyzing polarimetric observations might prevent the detection of grains with sizes of the order of or larger than the wavelength of the incident light. That could be the case with the maximum grain size of about 1 $\mu$m obtained by Canovas et al. (2013) to reproduce the polarized images of HD 142527 at the $H$ and $K_s$ bands.

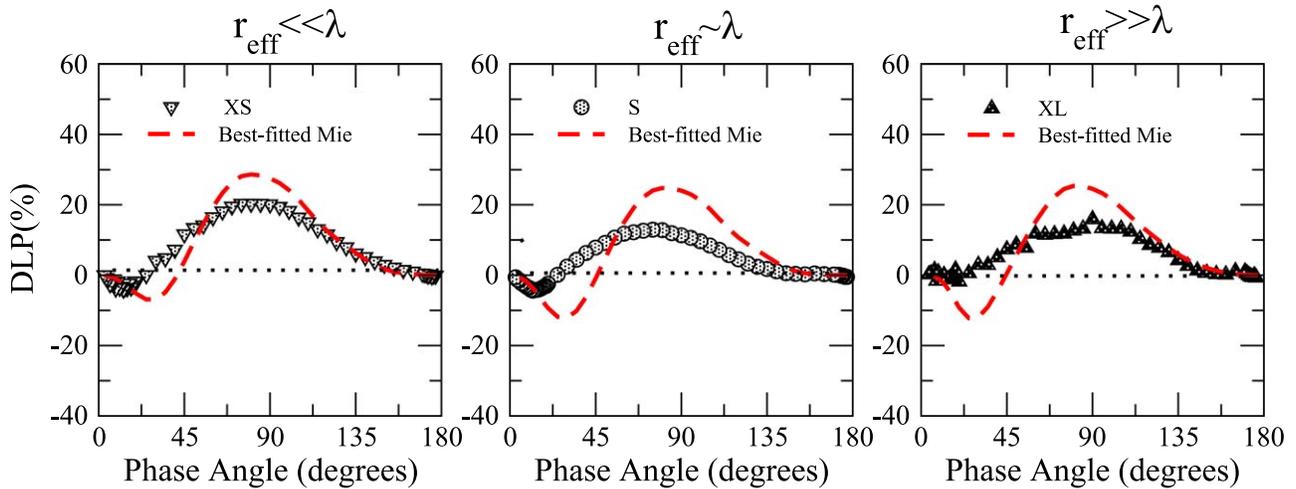

**Figure 10.** Experimental DLP curves for samples XS (left), S (middle), and XL (right). Red dashed lines correspond to Mie computations for the best-fit size distribution and refractive index parameters.

Table 3
Parameter Ranges in the DLP Fitting Procedure Considering Spherical Shapes

| | $q$ | $r_{min}$ ($\mu$m) | $r_{min}$ ($\mu$m) | $x_{eff}$ | $n$ | $k$ |
|---|---|---|---|---|---|---|
| **Fitting Range** | $2.5 \leqslant r \leqslant 4.5$ | $1E\text{-}3 \leqslant r \leqslant 10$ | $1 \leqslant r \leqslant 300$ | ⋯ | $1.1 \leqslant n \leqslant 4.0$ | $1E\text{-}5 \leqslant k \leqslant 10$ |
| Best-fit XS | 4.59 | 2.6E-2 | 0.22 | 0.63 | 2.09 | 2.96E-2 |
| Actual XS values | ⋯ | 0.14 | 0.8 | 4 | 1.65 | 1E-5 |
| Best-fit S | 4.49 | 7.3E-04 | 0.21 | 0.02 | 2.28 | 3.08E-3 |
| Actual S values | ⋯ | 0.45 | 10.71 | 17 | 1.65 | 1E-5 |
| Best-fit XL | 4.77 | 6.7E-2 | 0.23 | 0.11 | 2.13 | 7.7E-2 |
| Actual XL values | ⋯ | 18.6 | 112.2 | 575 | 1.65 | 1 E-5 |

**Note.** Best-fit parameters for samples XS, S, and L are compared to the actual parameter values of the corresponding sample.

### 6.1. Characterizing Size Grains from the Analysis of the Scattered Phase Function

The analysis of the forward scattering lobe is a widely used technique for retrieving the size distribution of small solid particles (see, e.g., Stojanovic & Markovic 2012; Gómez Martín et al. 2020). For smaller particles, the forward lobe spans a broader range of phase angles. Particles in the Rayleigh domain represent the limiting case with a symmetrical curve about its minimum located at 90°. As shown in Figure 5, the samples consisting of smaller particles (XS, S, and M) show a broader forward lobe. As the particle size increases (samples L and XL), the forward lobe becomes narrower. In the limiting geometric optics domain ($r \gg \lambda$), the diffraction peak is constrained into a narrow lobe around the exact forward direction. In remote sensing observations, the forward lobe is not always within the observable range of angles. In that case, the slope of the phase function at the side- and back-scattering angles can also be used to discriminate between various size regimes. As shown in Table 1, the $f$ estimator of the flattening of the phase function at side-scattering angles $(F_{11}(45°)/F_{11}(90°))$ increases with size, i.e., the larger the particles, the flatter the phase function at side-scattering angles. In the extreme case of particles significantly larger than the wavelength of the incident light, the phase function shows a slope opposite to the particles in the resonance regime i.e., the phase function for large pebbles is a strictly decreasing function of the phase angle in the measured back- and side-scattering angle range ($10° \leqslant \alpha \leqslant 140°$). Therefore, aside from the width of the forward-diffraction peak, which is strongly dependent on the grain size, the behavior of the phase function at the side- and back-scattering regions also provides information on the size regime of the scattering dust particles.

### 6.2. Characterizing Size Grains from the Analysis of the Degree of Linear Polarization

The study of the NPB and maximum of the DLP is an extended technique for characterizing dust particles in astronomical bodies like cometary coma (e.g., Lumme & Muinonen 1993; Kolokolova et al. 2007; Moreno et al. 2007; Hadamcik & Levasseur-Regourd 2009; Zubko et al. 2013b, 2016; Kiselev et al. 2015; Muinonen et al. 2015; Nezic et al. 2021) or protoplanetary and debris disks (Bastien & Menard 1988; Fischer et al. 1994; Kataoka et al. 2015, 2017; Kirchslager & Wolf 2014; Canovas et al. 2015, 2016; Esposito et al. 2018; Ohashi et al. 2018; Ren et al. 2019). We note that the NPB is known as polarization reversal within the stellar disks community. Both terms denote the same phenomenon: at small phase angles, the parallel component of the scattering field relative to the scattering plane (the plane defined by the observer, the star, and the dust cloud) is higher than the perpendicular component, so the DLP is negative. Specific features of the observed NPB such as its minimum value (DLP$_{min}$, $\alpha_{min}$) and the inversion angle ($\alpha_0$) are also used to discriminate among different asteroidal taxonomic classes (Penttila et al. 2005) and as diagnostic for their geometric albedo (Cellino et al. 2015). In this work we have shown that the position and magnitude of the NPB are strongly dependent on particle size. Dust particles with size parameters from ∼6 to ∼20 seem to be responsible for the NPB. As the relative amount of particles in that size range decreases, the NPB becomes shallower, disappearing within the error bars for sample XL. That is in agreement with Escobar-Cerezo et al. (2018), who found that by removing particles with size parameters below ∼12 from a lunar dust analog sample the NPB almost disappeared. Previous experimental DLP curves of cometary (Muñoz et al. 2000; Volten et al. 2006; Frattin et al. 2019) and Martian dust (Dabrowska et al. 2015) analogs also present the deepest NPB for those samples with a major contribution from particles with radii around 1 $\mu$m ($12 \lesssim x \lesssim 14$). In those cases, the broad size distributions of the samples prevented a link between the NPB and the mentioned size range to be concluded. The inversion angle, $\alpha_0$, is also dependent on the size regime. It spans from 25° for the samples consisting of smaller particles (XS and S) to 15° for sample L. For the limiting case of $r \gg \lambda$ (the quartzite pebble) the inversion angle increases again up to 30°, showing a shallow NPB with $DLP_{min} = -0.8\%$. Constructive interference of light multiply scattered by wavelength-scale surface or internal structure in the so-called coherent back-scattering mechanism (Shkuratov 1989; Muinonen 1990; Muinonen et al. 2015) could be responsible for the shallow NPB shown by large dust particles. Previous computations with Gaussian random shapes suggest a dependence of the width and magnitude of the NPB on size and refractive index (Muinonen et al. 2007).

As shown in Figure 6 and Table 2, the magnitude and position of the maximum of the DLP curve ($DLP_{max}$, $\alpha_{max}$) are strongly dependent on the particle size regime. For very small particles (sizes smaller than or of the order of the wavelength) the maximum polarization decreases with the size parameter. In the strict Rayleigh regime, the scattered light is 100% polarized at $\alpha = 90°$ when the incident light is unpolarized (Hovenier et al. 2004). The measured polarization curves for samples XS, S, M, and L show the decrease of $DLP_{max}$ with the size parameter. As the size parameter increases into the geometric optics domain (sample XL), the tendency of a decrease in $DLP_{max}$ with the size parameter is reversed. Further, the position of $DLP_{max}$ spans from 75° for the samples consisting of smaller particles to 140° for the pebble. The shift of $DLP_{max}$ toward larger phase angles with increasing size is in agreement with the computed results for Gaussian random spheres in the geometric optics regime (Liu et al. 2015; Escobar-Cerezo et al. 2017). A large $\alpha_{max}$ of about 140° is also observed in the $K_1$-band images of the debris disk HR 4796A obtained in the polarimetric mode of the Gemini Planet Imager (Perrin et al. 2015; Arriaga et al. 2020). This could indicate the presence of particles significantly larger than the wavelength in the HR 4796A disk.

As mentioned, a similar study using narrow size distributions of absorbing particles is being planned. This is of utmost importance to disentangle the effect of size and composition on the DLP curve.

### 6.3. Characterizing Size Grains from the Analysis of the Depolarization Ratio

The linear depolarization ratio at the exact backward direction ($\delta_L(0)$) is commonly used in lidar and radar remote sensing to characterize the physical properties of terrestrial atmospheric aerosols (Sassen 2005). Our experimental data show a strong dependence on particle size. However, as shown in Figure 7 (middle panel), the $\delta_L$ for significantly different size ranges tend to converge at 0°. The value of $\delta_L$ at the DS region ($8° \lesssim \alpha \lesssim 14°$) appears to be a better diagnostic of the size of atmospheric aerosols. Further, the $F_{22}(\alpha)/F_{11}(\alpha)$ ratio can be used as an indication of the nonsphericity of the dust particles because for spherical particles it is equal to 1 at all phase angles. For irregular dust particles, it differs from unity at all measured scattering angles.

When combining the $-F_{12}(\alpha)/F_{11}(\alpha)$ and $F_{22}(\alpha)/F_{11}(\alpha)$ curve, we find for sample S an anticorrelation between the NPB and the depolarization surge at the backward direction. Previously computed DLP and $F_{22}(\alpha)/F_{11}(\alpha)$ curves for coated Gaussian particles, coated aggregates of spheres, and vesicular particles with similar sizes also show a link between the NPB and the shape of the $F_{22}(\alpha)/F_{11}(\alpha)$ at the backward direction (Lindqvist et al. 2009; Lindqvist et al. 2011). This suggests a link between the scattering mechanism responsible for the NPB and the depolarization surge in the backward direction.

This work has been funded by the Spanish State Research Agency and Junta de Andalucía through grants LEONIDAS (RTI2018-095330-B-100) and P18-RT-1854, and the Center of Excellence Severo Ochoa award to the Instituto de Astrofísica de Andalucía (SEV-2017-0709). T.J. acknowledges the European Science Foundation (ESF) and the Ramon y Cajal Program of MICINN for their financial support.

### Data Availability

The experimental and synthetic scattering-matrix elements as functions of the scattering angle, size distribution tables, and SEM images are freely available at the Granada–Amsterdam light-scattering database (www.iaa.es/scattering) under the request of citation of this paper and Muñoz et al. (2012).

### ORCID iDs

O. Muñoz https://orcid.org/0000-0002-5138-3932
J. C. Gómez-Martín https://orcid.org/0000-0001-7972-085X
F. Moreno https://orcid.org/0000-0003-0670-356X
D. Guirado https://orcid.org/0000-0002-9228-1035
A. C. Caballero https://orcid.org/0000-0002-0571-6302
J. Milli https://orcid.org/0000-0001-9325-2511
F. Ménard https://orcid.org/0000-0002-1637-7393